# AUTOMATED DIAGNOSIS OF LUNG DISEASES USING VISION TRANSFORMER: A COMPARATIVE STUDY ON CHEST X-RAY CLASSIFICATION

**Muhammad Ahmad [1], Sardar Usman [2], Ildar Batyrshin [1], Muhammad Muzammil [3], K. Sajid [4], M. Hasnain [5], Muhammad Jalal [3] and Grigori Sidorov[1,*]**

[1]Centro de Investigación en Computación, Instituto Politécnico Nacional (CIC-IPN), Av. Juan de Dios Bátiz, Esq. Miguel Othón de Mendizábal S/N, Gustavo A. Madero, 07738 México City, México.
[2]Department of computer Science Grand Asian University of Sialkot, Pakistan
[3]Department of computer Science the Islamia University of Bahawalpur, Pakistan
[4]College of Computer Science and Technology, Zhejiang Normal University, Jinhua (321004), China.
[5]Department of Computer Science, Leads University Lahore.

**\*Corresponding Author:** Grigori Sidorov
Email: sidorov@cic.ipn.mx

**ABSTRACT**

**Background:** Lung disease is a significant health issue, particularly in children and elderly individuals. It often results from lung infections and is one of the leading causes of mortality in children. Globally, lung-related diseases claim many lives each year, making early and accurate diagnoses crucial. Radiographs are valuable tools for the diagnosis of such conditions. The most prevalent lung diseases, including pneumonia, asthma, allergies, chronic obstructive pulmonary disease (COPD), bronchitis, emphysema, and lung cancer, represent significant public health challenges. Early prediction of these conditions is critical, as it allows for the identification of risk factors and implementation of preventive measures to reduce the likelihood of disease onset

**Methods:** In this study, we utilized a dataset comprising 3,475 chest X-ray images sourced from from Mendeley Data provided by Talukder, M. A. (2023) [14], categorized into three classes: normal, lung opacity, and pneumonia. We applied five pre-trained deep learning models, including CNN, ResNet50, DenseNet, CheXNet, and U-Net, as well as two transfer learning algorithms such as Vision Transformer (ViT) and Shifted Window (Swin) to classify these images. This approach aims to address diagnostic issues in lung abnormalities by reducing reliance on human intervention through automated classification systems. Our analysis was conducted in both binary and multiclass settings.

**Results**: In the binary classification, we focused on distinguishing between normal and viral pneumonia cases, whereas in the multi-class classification, all three classes (normal, lung opacity, and viral pneumonia) were included. Our proposed methodology (ViT) achieved remarkable performance, with accuracy rates of 99% for binary classification and 95.25% for multiclass classification.

**INTRODUCTION**
Pneumonia [1] is a major health concern for both children and the elderly, stemming from lung infections, and is a major cause of death among children worldwide. Annually, approximately 1.4 million children die from pneumonia, accounting for 18% of all deaths among children under five-years old. Furthermore, pneumonia affects approximately two billion individuals globally each year [2]. Severe respiratory infections are prevalent and present an elevated risk for both young children and elderly individuals. Given the severity of this disease, extensive clinical diagnostic procedures are essential. One of the clinically validated methods for diagnosing pneumonia is chest radiography (radiographs) [3]. In traditional diagnostic practice, lung infections are assessed by a radiologist who then prescribes the necessary treatment based on their evaluation. This method, which relies on radiographs, is both time-consuming and susceptible to interpretative variability, which can influence the diagnostic accuracy. In addition, the severity of the disease may not always be clearly visible on radiographs, potentially prolonging the diagnostic process.

Computer-Aided Diagnosis (CAD) has recently emerged as a prominent subject in machine-learning research. CAD systems have demonstrated great benefits in the medical field, particularly for diagnosing breast cancer, interpreting mammograms, and identifying lung nodules. When using machine learning in medical imaging, significant features must be extracted. As a result, early algorithms frequently used handcrafted features to create CAD systems for image analysis [4, 5, 6]. Despite their historical use, handcrafted features have restrictions that vary by task and frequently fail to provide adequately robust functionality. Deep Learning (DL) models [26-32], notably Convolutional Neural Networks (CNNs), have shown greater ability to extract meaningful features for image categorization [7, 8].

Deep learning, a subset of artificial intelligence [33-40], has been shown to be extremely effective in identifying and diagnosing pneumonia using medical images such as chest X-rays [9]. Recent research has also highlighted the potential of deep learning techniques to solve issues associated with pneumonia diagnosis [10, 11, 12]. These algorithms have demonstrated great promise not only in research, but also in clinical practice. For example, an algorithm created to detect four different thoracic disorders using frontal chest radiographs was tested in an emergency medicine scenario and was found to enhance radiology residents' sensitivity [13].

In this study, we utilized a dataset comprising 3,475 chest X-ray images sourced from Mendeley Data provided by Talukder, M. A. (2023) [14], categorized into three classes: normal (n= 1250), lung opacity (n= 1125), and Viral pneumonia (1100). We applied five pre-trained deep learning models, including CNN, ResNet50[15], DenseNet[16], CheXNet [17], and U-Net, as well as two transfer learning algorithms such as Vision Transformer (ViT) and Shifted Window (Swin) to classify these images. This approach aims to address diagnostic issues in lung abnormalities by reducing reliance on human intervention through automated classification systems. Our analysis was conducted in both binary and multi-class settings. In the binary classification, we focused on distinguishing between normal and viral pneumonia cases, while in the multi-class classification, all three classes normal, lung opacity, and viral pneumonia were included. Our proposed methodology (ViT) achieved remarkable performance, with accuracy rates of 98.93% in binary classification and 95.25% in multi-class classification. Additionally, the research seeks to compare the results of these advanced transfer learning techniques with existing approaches.

The Study makes the following Contribution:

1. Propose, implement, and evaluate a transfer learning model designed to optimize the analysis of chest X-ray (CXR) images to enhance diagnostic accuracy and efficiency and improve diagnostic issues by reducing the reliance on human intervention through automated classification systems.
2. The tool was trained and tested using approximately 3,475 chest X-ray (CXR) images with binary and multi class sourced from Mendeley data provided by Talukder, M. A. (2023), including diverse pathological conditions such as Viral Pneumonia, Normal and Lung Opacity.

3. Conduct a comprehensive analysis and performance evaluation using various deep learning and transfer learning techniques, along with comprehensive visualizations.

The remainder of this paper is organized as follows: Section II outlines the literature survey and Section III Methodology and Design. Section IV: Results and Analysis. Section V discusses the limitations of the proposed methodology. Finally, Section VI presents the conclusions of the study.

**LITERATURE SURVEY**

Rahimzadeh, Mohammad et al. [18] trained deep convolutional networks to classify X-ray images into normal, pneumonia, and COVID-19 categories using two datasets, including 180 COVID-19 cases. They introduced a novel architecture combining Xception and ResNet50V2 to improve feature extraction. Tested on 11,302 images, their model achieved 99.50% accuracy for COVID-19 detection and 91.4% overall accuracy.

Rajpurkar, Pranav, et al. [19] developed CheXNet, a 121-layer convolutional neural network that detects pneumonia from chest X-rays with higher accuracy than practicing radiologists. Trained on the ChestX-ray14 dataset, which includes over 100,000 images and 14 diseases, CheXNet was evaluated against radiologist annotations and outperformed them in the F1 metric. The model also achieved state-of-the-art results for all 14 diseases in the dataset.

Janizek, Joseph D., et al. [20] tackle the issue of performance degradation in deep learning models due to dataset shift across different hospital systems. They use adversarial optimization to create models that are invariant to confounding variables, like view position in chest radiographs, improving generalization to external data. Their method enhances predictive accuracy and helps identify models that rely on confounders.

Gabruseva, Tatiana et al. [21] address the global impact of pneumonia and the challenges in detecting it via chest X-rays. They developed a computational approach using single-shot detectors, squeeze-and-excitation networks, augmentations, and multi-task learning to improve diagnostic accuracy. Their method was tested in the Radiological Society of North America's Pneumonia Detection Challenge, achieving one of the top results.

Farag, Abdullah Tarek, et al. [22] present Multi CheXNet, a multi task learning model for diagnosing, segmenting, and localizing pneumonia-like diseases from X-ray datasets. The model features a common encoder for efficient feature extraction and specialized decoders for task-specific details. It employs teacher forcing and transfer learning to improve performance on unseen diseases like COVID-19. Tested on various datasets, Multi CheXNet outperforms baseline models.

Ali, Wardah, et al. [23] introduce a framework for pneumonia detection that addresses class imbalance using Deep Convolutional GAN (DCGAN) and Wasserstein GAN with gradient penalty (WGAN-GP) for augmentation, along with Random Under-Sampling (RUS). Validated with the ChestX-Ray8 dataset, it achieves superior results using transfer learning on ResNet-50, Xception, and VGG-16 models.

This Study focuses on enhancing the accuracy and efficiency of pneumonia diagnosis from chest X-ray images by utilizing advanced deep learning methods, including CNNs and transfer learning models such as CheXNet and CXR-ViT. The goal is to overcome the shortcomings of traditional diagnostic techniques, decrease dependence on expert radiologists—particularly in underserved regions and develop a reliable automated system that improves early detection and treatment, thereby promoting better patient outcomes. The summary of prior studies related to Lung X-Ray image dataset detection is presented in Table 1.

**Table 1.** Prior studies related to Lung X-Ray image dataset.

| References | Model | Binary Class | Multi Class | Dataset | Binary Accuracy | Multi Accuracy |
| --- | --- | --- | --- | --- | --- | --- |

| | | | | Lung X-Ray image | | |
|---|---|---|---|---|---|---|
| Liu, X et al (2024) [24] | MobileNet-Lung | no | yes | dataset | no | 93.3 |
| Karla, Raghuram et al (2024) [25] | ResNet18 | no | yes | Lung X-Ray image dataset | no | 96.2 |
| Proposed | ViT | Yes | Yes | Lung X-Ray image dataset | 98.93 | 95.25 |

## METHODOLOGY AND DESIGN

The proposed methodology comprises four primary phases: Dataset Description, data preprocessing and augmentation, Training and testing phase and model evaluation.

### 1. Dataset preparation

In this paper, we utilized a dataset comprising 3,475 chest X-ray images sourced from Mendeley Data provided by Talukder, M. A. (2023) [14], categorized into three classes: normal (n=1250), lung opacity (n=1125), and viral pneumonia (n=1100). Further we divided dataset to binary class and multi class. In binary classification we categories our dataset to Normal and Viral Pneumonia and in multi class we took all the dataset. Each category links to a specific medical disorder, and all the images are labeled accordingly. The Corpus was divided into a training set and a validation set, with 80% of the data allocated for training and remaining 20% used for validation. This distribution of corpus enables the model to learn efficiently. Figure 1 depicts the distribution of labels for both the binary and multiclass classifications.

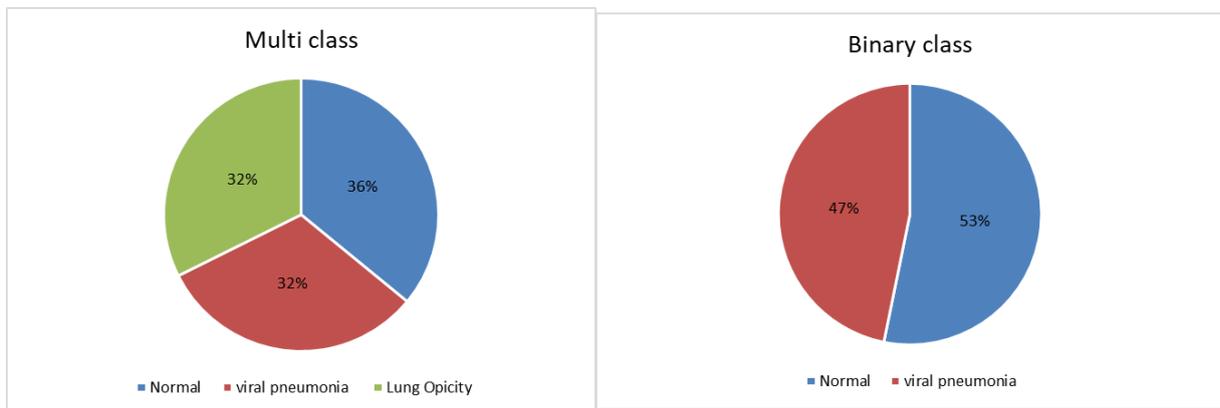

**Figure 1.** Label distribution entire dataset.

### 2. Data Preprocessing and Augmentation

Data preprocessing plays a crucial role in deep learning and transfer learning approaches. Often, real-world datasets are accompanied by noise, missing entries, duplicate values, different languages, and inconsistent formats. The preprocessing phase include variety of procedures to address these issues and ensure that the data are consistent and free from noise that could lead to incorrect results. During the data preprocessing and augmentation phase, we first resized the chest X-ray images to 224x224 pixels to match the input size required by the deep learning and Transfer learning models. To enhance the performance of the model's and avoid over fitting, numerous data augmentation techniques were applied during training such as random rotations, random resized cropping, and horizontal flipping of the images. Moreover, all images were normalized using a mean and standard deviation of 0.5, scaling the pixel values to fall between -1 and 1. This step ensured that the chest X-ray images were used to improve the performance of the models during training and testing phase.

## 3. Training and Testing phase

We employed five pre-trained deep learning models, including CNN, ResNet50, DenseNet, CheXNet, and U-Net, as well as two transfer learning algorithms such as Vision Transformer (ViT) and Shifted Window (Swin) to classify and identify the lung abnormalities based on chest X-ray images. We selected Pre-train deep learning and transformer which has the best ability to capture spatial hierarchies in the images. The presented models trained on 80% data and is tested on remaining 20% as shown in figure 2.

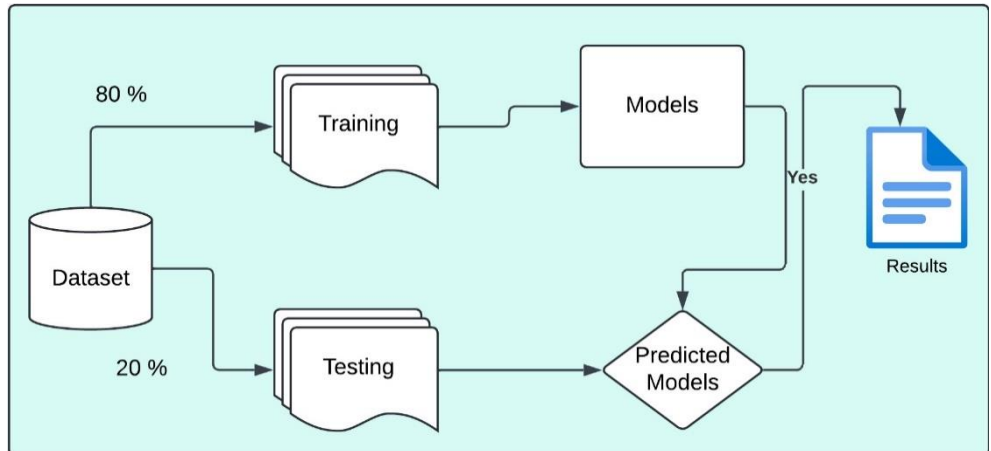

**Figure 2.** Training and testing phase.

## 4. Model Evaluation

In the model evaluation phase, we utilized key metrics, such as accuracy, precision, recall, and F1-score, (as seen in equations 1, 2, 3 and 4) to evaluate the performance of our proposed model. Through a thorough examination of these metrics, we gained invaluable insights into the capacity of the models to accurately classify chest X-ray images, ultimately providing insightful feedback on their effectiveness in identifying lungs abnormalities within chest X-ray images.

To enhance the performance of the model multiple epochs, batch sizes, and the learning rate were employed and best epoch, batch size and learning rates are chosen. Finally, the top-performing models were selected based on their accuracy, precision, recall and f1-score across different lung disease categories.

$$\text{Precision} = \frac{\text{True Positives}}{\text{True Positives} + \text{False Positives}} \quad (1)$$

$$\text{Recall} = \frac{\text{True Positives}}{\text{True Positives} + \text{False Negatives}} \quad (2)$$

$$\text{Accuracy} = \frac{\text{True Positives} + \text{True Negatives}}{\text{total Population}} \quad (3)$$

$$\text{F1 Score} = \frac{2 * (\text{Precision} * \text{Recall})}{\text{Precision} + \text{Recall}} \quad (4)$$

## RESULT AND DISCUSSION

This section will examine the outcomes of our predicted models. We employed deep learning and transfer learning-based approaches to identify the most effective solution for detecting lung abnormalities based on chest X-ray images.

### 1. Deep learning

Figure 3 and 4 displays the hyper-tuning parameters of the top-performing models for binary and Multi classification respectively. We systematically analyzed the influence of various epoch and batch sizes on our experimental results through ablation research in multi classification. Thus, we were able to determine the specific contributions of each parameter to attain the optimal performance of the system. All parameters were chosen based on evaluations with different parameter values.

### 1.1 Binary classification

In figure 3 bar chart presents the hyper-tuning parameters of the top-performing models for binary classification results of five pre-trained deep learning models such as: ResNet50, DenseNet, CNN, Unet and CheXNet—on a binary image classification task using a lung disease dataset to identify lung abnormalities. ResNet50 achieved the highest accuracy, scoring 97.87%, followed closely by DenseNet with a score of 97.50%. The CNN model, however, performed significantly lower, with an accuracy of 97.26%. CheXNet, another deep learning model designed specifically for medical imaging, scored 97.45%. The results show that ResNet50 and Unet are highly effective for this binary classification task, likely due to their deep architectures that excel in extracting complex features from medical images. This comparison highlights the importance of selecting pre-trained deep learning models that can handle the specific complexities of medical image data.

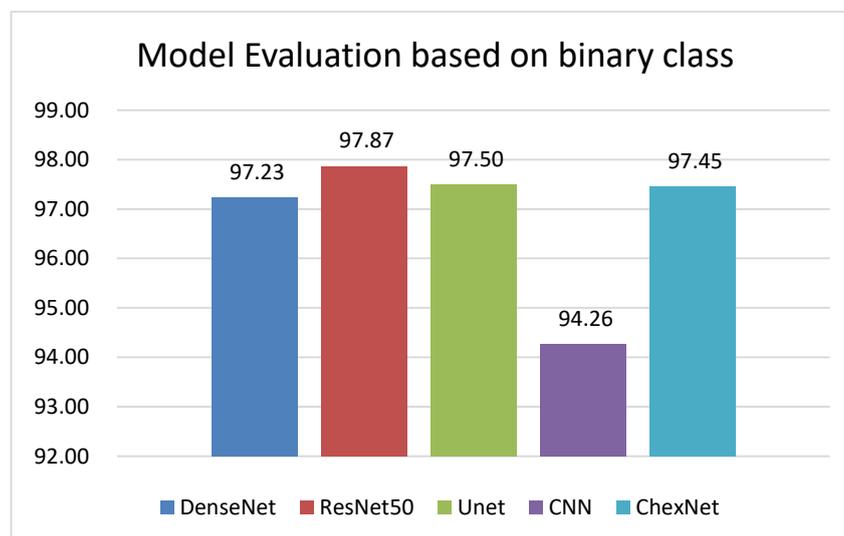

**Figure 3.** Deep learning models for binary class.

### 1.2 Multi classification

Figure 4 bar chart illustrates the accuracy results of different deep learning models. Among the models, DenseNet achieved the highest accuracy at 93.24%, closely similar level of effectiveness followed by CheXNet with 93.24%. The Unet model, however, lagged behind with a significantly lower accuracy of 87.63%. These results suggest that more advanced architectures like ResNet50 and DenseNet are better suited for handling the complexity of multi-class classification in this lung disease dataset.

Page | 6

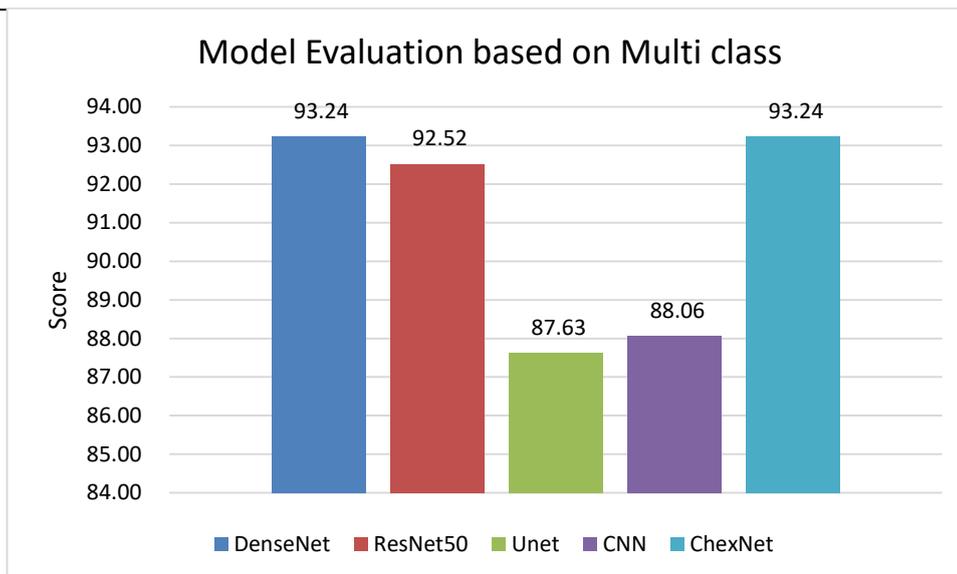

**Figure 4.** Deep learning models for multi class.

2. **Transformer Results**

    **2.1 Binary classification**

Figure 5 shows the comparison of model evaluation results of two different transfer learning models such as: ViT and SWIN—on a binary classification task. The ViT model achieved the highest accuracy at 98.93%, followed by the SWIN model, and the SWIN model with the lowest accuracy at 98.30%. This indicates that the ViT model performed best on this dataset, suggesting it is the most effective among the other for this specific lung disease classification task.

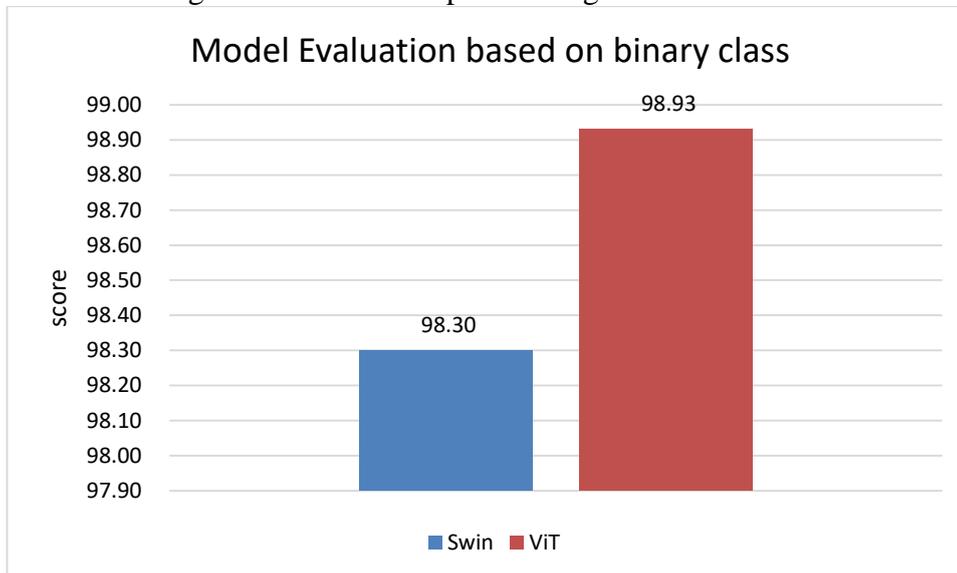

**Figure 5.** Transformer for binary class.

   **2.2 Multi classification**

In figure 6 bar chart illustrates the comparative accuracy of the models. The ViT model stands out with the highest accuracy at 95.25%, indicating its strong performance. The SWIN model follows closely demonstrating a similar level of effectiveness. However, the SWIN model shows a notable drop in accuracy, scoring 94.40%, which suggests it may be less suited for this particular multi-class classification task compared to the other model.

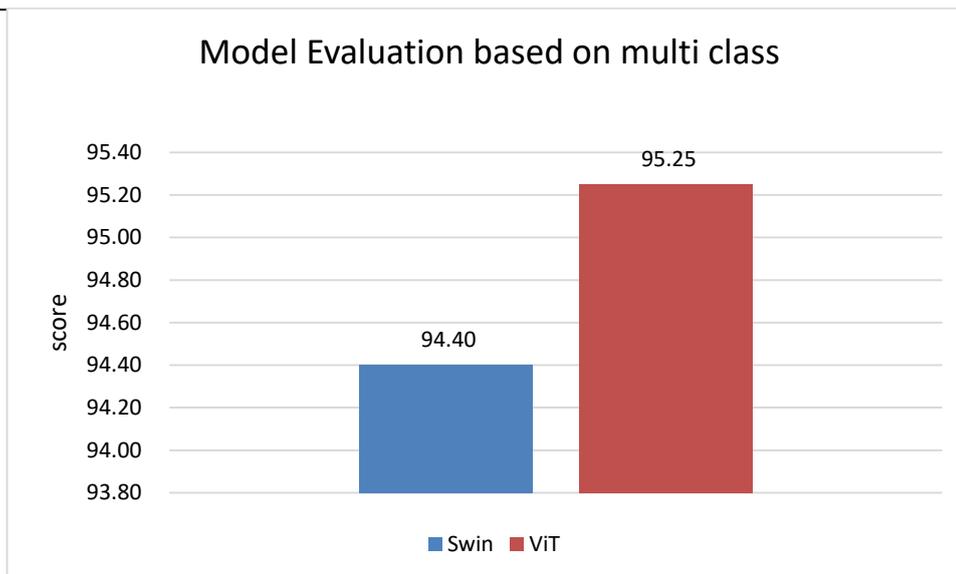

**Figure 6.** Transformer for multi class.

Table 2 displays the optimal fine-tuning parameters for a pre-trained Vision Transformer (ViT) model used for both binary and multi-class classification tasks. The best hyper parameters were identified through a grid search, with the following parameter ranges: learning rates of 1e-5, 1e-2, 2e-5, 3e-5, and 3e-4; epochs of 3, 9, 20 and 25; batch sizes of 8, 32, 64, and 128; weight decay values from 0.01 to 0.1; dropout rates of 0.02 and 0.1; and warm-up steps between 0.03 and 0.1. These parameters were chosen to optimize training efficiency and ensure robust performance of the model across various classification scenarios.

**Table 2.** Optimum Values Identified for the Hyper-Parameters of ViT model.

| Hyper parameter | Grid search |
| --- | --- |
| Learning rate | 1e-5,1e-2, 2e-5, 3e-5, 3e-4 |
| Epoch | 3, 9, 20, 25 |
| Batch size: | 8, 32, 64, 128 |
| Weight Decay | 0.01–0.1 |
| Hidden dropout | 0.02, 0.1 |
| Warm-up Steps | 0.03–0.1 |

3. **Error Analysis**

Table 3 presents the class-wise scores, while Figure 7 illustrates the confusion matrix for both binary and multi-class classification achieved by our proposed model. Notably, our model exhibited superior precision in the "Viral Pneumonia" class. While figure 8(a) and b shows the Confusion matrix of our proposed model in both binary and multi class. While figure 9 representing the training and validation performance metrics of our proposed model over multiple epochs in binary class while figure 9 depicts the training and validation performance of different epochs in multi classification task. The system was trained on 20 epochs at the initial stage the accuracy and validation loss were 0.9416 and 0.1723 in binary class and in multi class at initial stage 0.9281 and 0.3775 at the 20 epoch the accuracy and validation loss were 98.94 and 0.1056 in binary class while 0.9525 and 0.2937 respectively. These improvement clear shows that the model learn effectively from the training data and made more accurate prediction.

**Table 3.** Class wise score for the proposed methodology.

| Categories | Precision | Recall | F1-Score | Support | Accuracy |
|---|---|---|---|---|---|
| Binary Classification | | | | | |
| Normal | 0.99 | 0.99 | 0.99 | 250 | 99% |
| Viral Pneumonia | 0.99 | 0.99 | 0.99 | 220 | |
| Multi classification | | | | | |
| Lung Opacity | 0.97 | 0.91 | 0.94 | 225 | |
| Normal | 0.91 | 0.97 | 0.94 | 250 | 95.25% |
| Viral Pneumonia | 0.99 | 0.98 | 0.98 | 220 | |

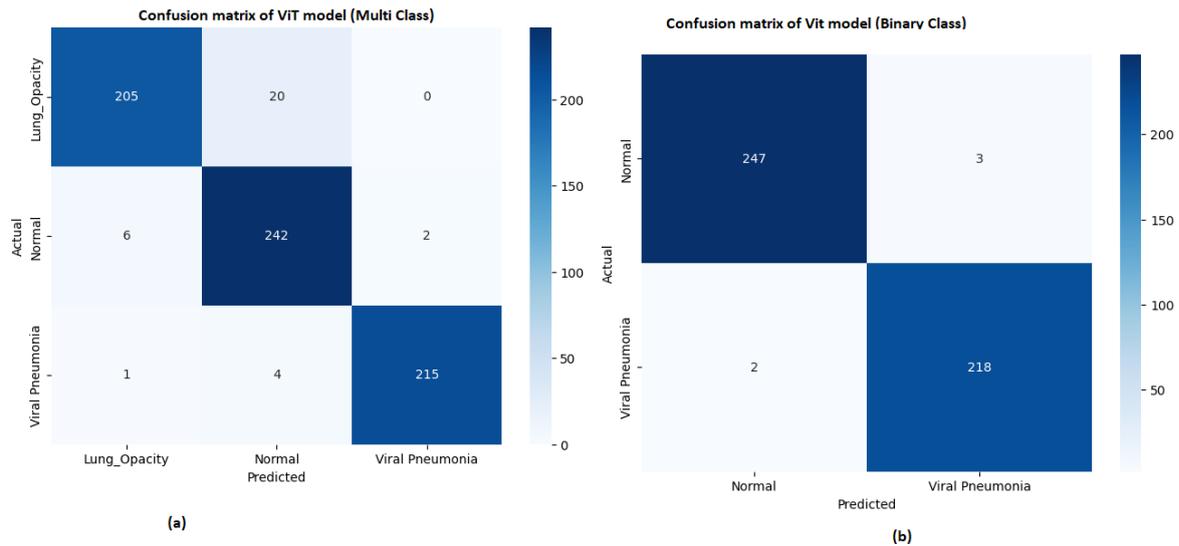

**Figure 7.** Confusion matrix of ViT in binary and multi class.

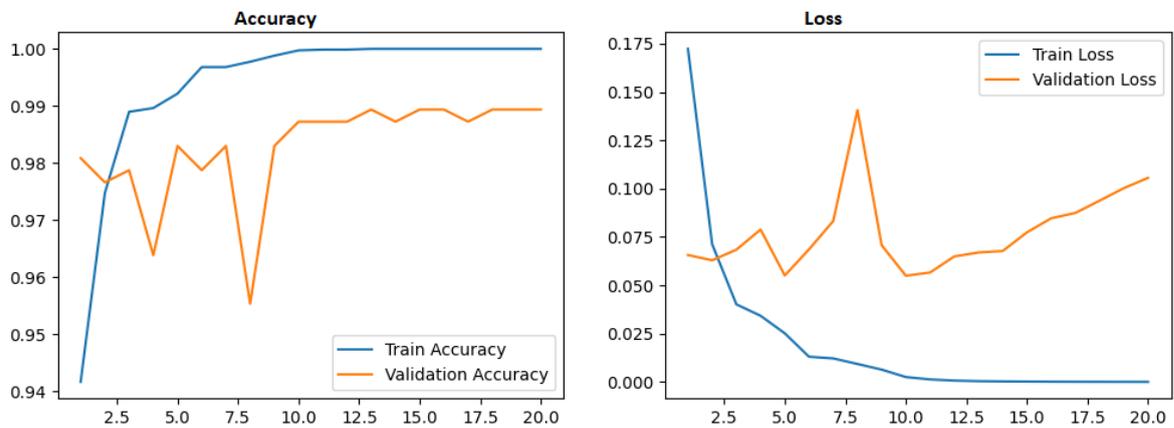

**Figure 8.** Training and validation performance of different epochs in binary class.

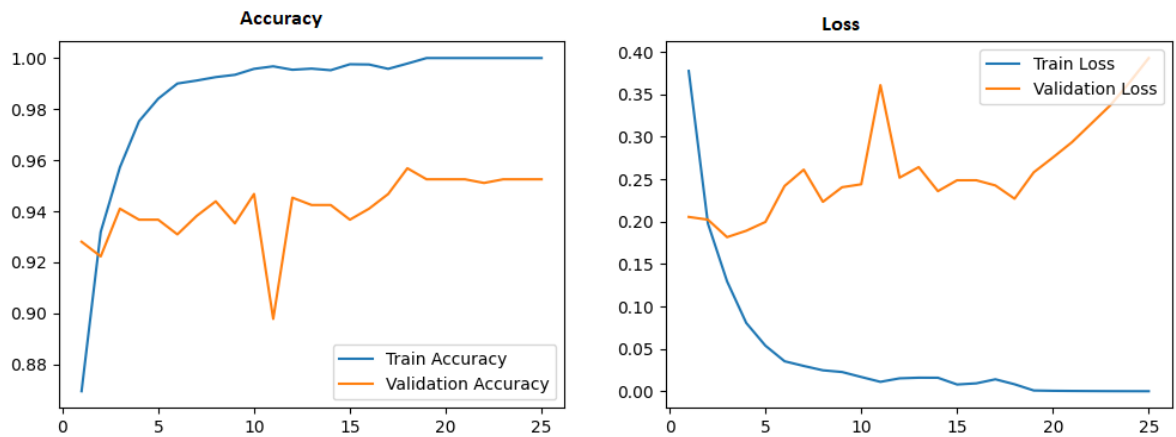
**Figure 9.** Training and validation performance of different epochs in multi class.

## LIMITATION OF PROPOSED SOLUTION
In acknowledging the limitations of this study, we recognize that the sample size may be insufficient for broad generalization of our predictions. However, the diverse range of attributes and features within our dataset provides a comprehensive representation of each patient, enhancing the robustness of our analysis despite the limited sample size.

## CONCLUSION
Our study examines Clinical Decision Support Systems (CDSSs) in healthcare. It underscores the importance of accurately identifying respiratory diseases, which are among the most critical public health issues today. By leveraging advanced Transfer learning models, the research demonstrates the potential for more precise feature extraction from chest X-ray images, leading to improved classification of these life-threatening conditions. Our proposed methodology (ViT) achieved remarkable performance, with accuracy rates of 98.93% in binary classification and 95.25% in multi-class. It addresses diagnostic challenges by minimizing the reliance on human intervention through advanced automated classification systems. This advancement not only enhances diagnostic accuracy but also supports more timely and appropriate treatment decisions, ultimately contributing to better patient outcomes. Future work could extend these methods to further refine detection and classification capabilities for other complex respiratory disorders.

## FUNDING
This research did not receive any funding.

## ACKNOWLEDGEMENTS
This work was done with partial support from the Mexican Government through the grant A1-S-47854 of CONACYT, Mexico, grants 20241816, 20241819, and 20240951 of the Secretaría de Investigación y Posgrado of the Instituto Politécnico Nacional, Mexico. The authors thank the CONACYT for the computing resources brought to them through the Plataforma de Aprendizaje Profundo para Tecnologías del Lenguaje of the Laboratorio de Supercómputo of the INAOE, Mexico and acknowledge support of Microsoft through the Microsoft Latin America PhD.